\documentstyle[12pt]{article}
\begin{document}
\baselineskip=0.6cm
\newcommand{\EQ}{\begin{equation}}
\newcommand{\EN}{\end{equation}}
\newcommand{\EQA}{\begin{eqnarray}}
\newcommand{\EQN}{\end{eqnarray}}
\newcommand{\e}{{\rm e}}
\newcommand{\Sp}{{\rm Sp}}
\renewcommand{\theequation}{\arabic{section}.\arabic{equation}}
\newcommand{\Tr}{{\rm Tr}}
\renewcommand{\thesection}{\arabic{section}.}
\renewcommand{\thesubsection}{\arabic{section}.\arabic{subsection}}
\makeatletter
\def\section{\@startsection{section}{1}{\z@}{-3.5ex plus -1ex minus
 -.2ex}{2.3ex plus .2ex}{\large}}
\def\subsection{\@startsection{subsection}{2}{\z@}{-3.25ex plus -1ex minus
 -.2ex}{1.5ex plus .2ex}{\normalsize\it}}
\def\appendix{
\par
\setcounter{section}{0}
\setcounter{subsection}{0}
\def\thesection{\Alph{section}}}
\makeatother
\def\thefootnote{\fnsymbol{footnote}}
\begin{flushright}
hep-th/0109058\\
UT-KOMABA/01-05\\
\end{flushright}
\vspace{1cm}
\begin{center}
\Large
World-Sheet String Duality \\and the Hidden Supersymmetry
\footnote{
Talk presented at the 10th International Symposium 
on String Theory (July 3-7, Fukuoka, Japan, 2001) 
and a part of lectures given at the 
Summer Conference on Strings  (July 16-27, Beijing, China, 2001). }

\vspace{1cm}
\normalsize
{\sc Tamiaki Yoneya}
\footnote{
e-mail address:\ \ {\tt tam@hep1.c.u-tokyo.ac.jp}}
\\
\vspace{0.3cm}

{\it Institute of Physics, University of Tokyo\\
Komaba, Meguro-ku, 153 Tokyo}

\vspace{1cm}
Abstract
\end{center}

 It is reviewed how space-time supersymmetry is realized 
nonlinearly in 
open superstring theory without making the GSO projection. 
We show that the world-sheet string dualities, 
viz. dualities of open-closed strings and of open-open 
  strings, play crucial roles for the 
existence of 10 dimensional $N=2$ supersymmetry in a
spontaneously broken phase. We also speculate on a  possible
mechanism of the  restoration of supersymmetry  from the
viewpoint of world-sheet  dynamics. 

\vspace{1cm}
\section{Introduction}
One of the interesting new concepts established in recent 
advances of superstring theory is the notion of unstable
D-branes \cite{sen0}.  In a sense, unstable D-branes may be
viewed as more fundamental than stable (BPS) D-branes, since 
stable D-branes can be classified using K-theory starting
from  the systems of unstable D9 (and/or anti-D9) branes in
type  IIA(B) theories \cite{witthora}. Recent investigations in
noncommutative  field theory and also in string field theory 
suggest further that all D-branes may be understood 
even dynamically on the basis of such unstable systems. 
However, most of such dynamical discussions have so far
been restricted to bosonic strings. In particular, 
the role of space-time supersymmetry (susy) 
in unstable systems has not been understood appropriately. 
The space-time supersymmetry is of course 
broken in such unstable vacua. Nonetheless its existence 
in a hidden (spontaneously broken) form must be 
important, since the spontaneously broken 
supersymmetry  can  in principle  constrain the 
dynamics in some definite manner.

In the present talk, I would like to review,   
on the basis of two previous works \cite{yo}\cite{harayo}, the
status of the above question. 
Throughout the discussions below, 
I will mostly emphasize the standpoint of world-sheet 
dynamics, although string field theory 
is in general potentially suitable 
for investigating spontaneously broken 
symmetries. The main reason for
this is that  the world-sheet string dualities turn out to be 
crucial for the consistent implementation of hidden susy 
 for nonzero $\alpha'$, while 
string field theory is  not, unfortunately 
at least in its present 
level of development,  convenient 
for exhibiting such stringy characteristics.

\section{Evidence of hidden supersymmetry}
 
Questions which we should ask first may be those: How 
 can we see the spontaneouly broken susy in terms of 
perturbative theory of the open strings describing  unstable
D-branes?  Are there any concrete signatures of the broken
susy?  A possible clue in  answering such questions is that
unstable D-branes  should be coupled to gravity or closed
strings  consistently at least in the sense of perturbation 
theory. One among the most intriguing phenomena in string
theory is that open-string  loop diagrams are dual to
closed-string tree diagrams.  This implies that open-string
loop  amplitudes themselves without direct coupling 
to closed strings must
exhibit certain trace of the type II supersymmetry of closed
strings, to the  extent that they are acceptable as
perturbative amplitudes. 

Let us take
the  simplest example of a single D9-brane in 
type IIA theory and consider the 
one-loop partition function of a single open string 
in the NSR formalism, which is, ignoring the momentum part and 
fixing the moduli parameter $\tau$, 
\begin{equation}
Z(\tau)=\Tr_{NS}\Bigl(
q^{2N_{NS}-1}
\Bigr)
-\Tr_{R}\Bigl(
q^{2N_{R}}
\Bigr)
\label{zett}
\end{equation}
where $q=\e^{-\tau/2\alpha'}$ and 
$N_{NS}, N_{R}$ are the level operators  
for the NS sector and R sector, respectively. 
Note that we do not here make the standard GSO projection. 
Then the partition function has nonzero 
contribution only from the oppositely GSO projected sector 
as 
\EQ
Z(\tau)={1\over \prod_{n=1}^{\infty}(1-q^{2n})^8}
\Bigl({1\over 2q}
\prod_{n=1}^{\infty}(1+q^{2n-1})^8 +
{1\over 2q}\prod_{n=1}^{\infty}(1-q^{2n-1})^8
-{1\over 2}16 
\prod_{n=1}^{\infty}(1+q^{2n})^8
\Bigr) .
\label{partfunc}
\EN
In the case of the ordinary GSO projected sector, 
the sign in front of the second term inside the large round
brace is  opposite, and hence $Z(\tau)$ vanishes due to the
well known  Jacobi's `{\it abstruse}' identity. This vanishing
is a  consequence of the matching of the degrees of 
freedom between bosons and fermions at each fixed mass level. 
Conversely, the nonvanishing result of the partition function
(\ref{partfunc}) means that  there is no matching of the
degrees of freedom  between bosons and fermions at fixed mass
levels. 

However, if supersymmetry is hidden, there must be the similar 
matching with respect to 
the total degrees of freedom. Namely,
the number  of the degrees of freedom must be equal between 
bosons and fermions when we ignore the 
difference of masses. Furthermore, in the case of 
ordinary local field theories, it is well known
that spontaneously broken  supersymmetry usually leads to the
following  mass formula for the linearized excitation levels:
\EQ
\Tr\Bigl(
(-1)^F M^2 
\Bigr)=0 
\label{tracemsquare}
\EN
where $F$ is the {\it space-time} fermion number. 
In the present case, this is not well defined as it stands
since there are an infinite number  of states with increasing
masses. 
However,  the partition function can be used to obtain
a general (regularized)  
 mass formula by 
\EQ
A_n \equiv \lim_{\tau\rightarrow 0}
\Tr\Bigl(
(-1)^F M^{2n}\e^{-\tau M^2}
\Bigr) =(-1)^n\lim_{\tau\rightarrow 0}{d^n\over d\tau^n}
Z(\tau) .
\EN 
This implies that the 
whole information on the number of degrees of
freedom and on the 
mass formula of open-string spectrum as well can
equivalently be  expressed in terms of the low-energy spectrum 
of closed strings in the dual channel, 
since the short-time
limit $\tau 
\rightarrow 0$ 
in the open-string channel is equivalent 
to the large time
$1/\tau
\rightarrow \infty$ propagation of closed strings 
$(\sigma=\tau/2\pi\alpha')$:
\EQ
Z(\tau) \rightarrow 
{\sigma^4\over 2}\Bigl(16 + 16 + (256+ 256) \e^{-2\pi/\sigma} 
+ O(\e^{-4\pi/\sigma})\Bigr) 
\EN
which leads to 
\EQ
A_{n}=0  , \quad (n\ne 4). 
\label{traceformula}
\EN 
Thus, not only the total degrees of freedom 
match between (space-time) bosons and fermions, but also 
the quantities (mass)$^{2n}$ averaged over 
the whole open-string degrees of freedom are equal 
between bosons and fermions for all $n$ except for 
$n=4$. It is more than natural to  
imagine the existence of supersymmetry 
behind this surprising cancellations. 
Since the above duality is a consequence of the 
interplay of the whole tower of 
open string states, we expect that 
the exact hidden supersymmetry can only 
be properly explained by taking account of 
the whole massive higher excitations of open strings, 
at least for nonzero $\alpha'$. 
It should be emphasized that the consistency 
of the above interpretation is owing to the 
fact that the closed string channel is automatically 
restricted to the ordinary GSO projected sector. 
Otherwise, we would have tachyon in the closed-string 
channel which completely ruins the vanishing of $A_n$ 
because of the divergence in the limit $\tau \rightarrow 0$. 
In the open-string language,  
the GSO projection  for closed strings comes
about by the insertion of the factor $(-1)^F$ 
in the partition function.  This restricts the 
excitation of fermionic world-sheet fields 
to be odd in both left and right moving 
fields in the closed-string sector, irrespectively 
of NS-NS or of R-R sector.  This is nothing 
but the GSO projection of closed strings. 

Here we have treated the simplest case 
of a single D9-brane in type IIA theory. 
Extensions to multiple D9 branes and to 
D9-$\overline{{\rm D}9}$ in type IIB theory is 
straightforward, since the above open-closed 
string duality is valid separately 
for both the sectors of ordinary GSO and opposite GSO
projections.  Recall that the open strings stretching between 
D9 and anti-D9 brane are oppositely GSO projected. 
All we have to be careful is to 
avoid the Ramond-Ramond anomaly. For example, in the 
case of type IIB theory, we have to consider 
the same number of D9 and anti-D9 branes. This will 
ensure that the hidden susy for type IIB case 
is the  chiral $N=2$ susy. 

In the zero-slope limit
$\alpha'\rightarrow 0$,  we expect that the consistent
coupling  of massless open string states with supergravity 
fields is guaranteed again by the open-closed string 
duality. For the resulting low-energy effective theory to be 
consistent, 
the massless open-string sector must be  described by an
appropriate nonlinear  realization of $N=2$ supersymmetry. 
The latter can be  obtained by a systematic $\alpha'$
expansion  of the full theory,  
as we will discuss later. This amounts to eliminating 
all the massive modes, including the tachyon, in terms 
of massless modes using the equations of motion. 
Once we include the tachyon of the open-string sector,
however,  we must take the other massive excitations into
account  on equal footing with the tachyon. In principle, it
is possible to  derive some effective theory expressed  in
terms of massless open string states {\it and only} tachyon by
eliminating the other massive excitations through  the equation
of motion. However, it seems very   hard to recast such a
theory into closed and useful form.\footnote{ Such a
possibility was  touched upon
briefly in 
\cite{yo} to motivate discussions on the hidde susy 
and was further studied
more systematically in 
\cite{teraue}. }

Another interesting class of total spontaneous breaking of 
supersymmetry is obtained when the supersymmetry
is  broken by an orientifold plane, being accompanied 
by the suitable number (32) of $\overline{D9}$ branes 
\cite{sugimoto}. It was confirmed \cite{moriyama} 
that the above observation applies to this case as well. 
Including this case, an extensive discussion on the 
consistency (viz. anomaly cancellation) of various models with
spontaneous broken  supersymmetry was given in \cite{schwitt}.

\section{Massless Ramond fermion as Nambu-Goldstino} 

Let us next study another important trace of 
spontaneously broken supersymmetry. 
If we first switch off gravity 
and consider open strings propagating 
in flat space-time, there must be the Nambu-Goldstone 
state (Goldstino, for brevity) associated with the global
supersymmetry of the open-string sector.  
 In the case of type II theories, we should have  two
corresponding Goldstino excitations in 10 dimensions.  Such
Goldstinos would be `eaten' by  gravitinos  through super Higgs
effect when we take the mixing with closed-string  states into
account.  Nevertheless, in perturbation theory, it is
meaningful  and useful to study Goldstino as the crucial 
signature of spontaneous breaking of supersymmetry 
in the open-string sector.  Let us
therefore consider whether and how  the massless fermionic
ground states of the  open fermionic strings can be regarded
as the  Goldstino. 

In the field-theory language, the spontaneous
symmetry breaking  is in general
reflected to the appearance of inhomogeneous terms in the
corresponding symmetry  transformation law of the Goldstone 
field $\psi$, $$\delta \psi =
\epsilon + \mbox{linear and higher terms},$$ or equivalently,
to the presence of linear terms
$$j^{\mu}=\gamma^{\mu}\psi+\mbox{ quadratic and higher
terms},$$ in the corresponding conserved current. In the first
quantized  world-sheet picture, both of these features are 
summarized into the following property of the 
world-sheet (super)current: The conservation of the 
world-sheet charge is violated, but the violation 
is compensated by the emission or absorption of the 
Goldstino from the vacuum at zero momentum. 
More concretely, the divergence of  the 
world sheet (super)current must be proportional 
to the zero-momentum limit of the vertex operator 
of the massless Goldstino state. 
The violation of world-sheet charge 
conservation occurs because the first quantized
world-sheet supercurrent  only represents the effect
of the quadratic terms of the  space-time 
current.  On the other hand, the compensation of 
charge nonconservation due to
the  Goldstino emission at zero momentum 
corresponds to the shift  $\delta
\psi$ of the  Goldstino field, which is represented by 
insertion of the zero-momentum vertex operator 
and corresponds to the action generated by the linear term in
the  space-time charge associated to the linear term of 
the conserved space-time current.  
Actually, 
the massless gauge bosons, including graviton 
and its partners, can all be regarded as 
Goldstone excitations in similar manner, corresponding to
globally  nontrivial gauge transformations \footnote{
The simplest example of this is the $U(1)$ gauge 
transformation $\delta A_{\mu}(x)=\partial_{\mu}
\lambda(x)$ with 
$\lambda(x) =c_{\mu}x^{\mu}$ 
with $c_{\mu}$ being a constant vector.} which are 
broken in the ordinary perturbative vacua. 

It is easy to check that, without the 
ordinary GSO projection,  the world-sheet
supercurrent  indeed satisfies the above criteria. First, in
the  NSR formalism, the world-sheet current 
$s_{\alpha}^a  \quad (\partial_a s_{\alpha}^a=0)$ 
is given as
\[
s_{\alpha}^0(\tau,\sigma) = {1\over 2}(S_{\alpha}\e^{-\phi/2}(\tau-\sigma) 
+ S_{\alpha}\e^{-\phi/2}(\tau+\sigma)) ,
\]
 \[
s_{\alpha}^1(\tau,\sigma) = {1\over 2}(S_{\alpha}\e^{-\phi/2}(\tau-\sigma) 
- S_{\alpha}\e^{-\phi/2}(\tau+\sigma)) ,
\]
in terms of the spin operator $S_{\alpha}(\tau\pm \sigma)$ 
($\alpha$=32 component spinor index in 10 dimensions) 
and  bosonized ghost $\phi(\tau\pm \sigma)$.  
In the presence of oppositely GSO projected states, 
the conservation of the world-sheet charge \[
Q_{\alpha}(\tau)=\int_{0}^{\pi}d\sigma \, s_{\alpha}^0(\tau, \sigma)
\] 
is in general violated at the boundary of open strings:
\[
{d Q_{\alpha}(\tau)\over d\tau}
=-\int _0^{\pi}d\sigma \, \partial_1 s_{\alpha}^1(\tau,
\sigma)=-S_{\alpha}\e^{-\phi/2}(\tau-\pi)  +
S_{\alpha}\e^{-\phi/2}(\tau+\pi) 
\]
\[
\Downarrow
\]
\EQ
{d Q_{\alpha}(\tau)\over d\tau}=-
2S_{\alpha}\e^{-\phi/2}(\tau-\pi) ,
\label{chvio}
\EN
since the spin operator can be double-valued 
in the sense \footnote{
In the Euclidean metric on the world-sheet, this 
is $S(z)=-S(e^{i2\pi}z)$.} 
\[
S(\tau) =-S(\tau -2\pi).
\]
In the case of closed-string vertex operators, such 
double-valuedness is not acceptable,  since it 
introduces nonlocality in the world-sheet field theory. 
In the case of open strings, on the other hand, 
this is allowed, as long as the double-valued 
operators live only at the boundary. The right hand side 
of (\ref{chvio}) coincides with the vertex operator (of 
zero momentum) of the massless Ramond ground state. 
Thus, we can indeed identify the (singlet) 
Ramond ground state as the Goldstino state 
corresponding to the spontaneous broken 
$N=2$ susy: $N=2$ since there are 32(=16+16) components 
of the massless Goldstino fields. 

At a very formal level, the above structure can in fact be 
translated into the following transformation law 
of string fields in the framework of Witten's 
open superstring field theory:
\EQA
\delta a &=& \epsilon^{\alpha}Q_{\alpha} \psi , \\
\delta \psi &=&  
\epsilon^{\alpha}Q_{\alpha}Q_L{\cal I}+
X({\pi\over 2}) \epsilon^{\alpha} Q_{\alpha}a ,
\label{fermitrans}
\EQN
where $a$ and $\psi$ are the bosonic and fermionic parts, 
respectively of the string fields. Other notations are as 
follows:
$I$ = identity field, $Q_L$ = half-integrated BRST operator, 
$X({\pi\over 2})$ = picture changing operator inserted at the 
mid point $\sigma =\pi/2$. For more details, see \cite{yo}
and the  references therein. 
The first term of the right hand side of (\ref{fermitrans}) 
corresponds to the Goldstino nature of the string field.  
Unfortunately, however, it is not clear whether 
this transformation law can be justified rigorously, 
because of the well known problems related to  
the midpoint anomalies of superstring field theory. 
Berkovitz \cite{berko} proposed an interesting new formulation 
in which the mid-point insertion of picture changing 
operator is avoided. 
It is desirable to extend the above susy transformation 
law to this formulation. 

The same conclusion as above can be reached by using the 
Green-Schwarz formalism as well. To see this, it is 
useful to reconsider the partition function (\ref{partfunc}) 
again. Although this is nonvanishing, it can be 
rewritten by using the Jacobi identity as 
\[
{1\over q}{\prod_{n=1}^{\infty}(1-q^{2n-1})^8 \over 
\prod_{n=1}^{\infty}(1-q^{2n})^8} . 
\]
This result coincides precisely with the spectrum 
of the Green-Schwarz open string when we assume that the 
world-sheet spinor fields $S_{\alpha i} \, \, (i=1, 2)$
 (which
transform as  space-time spinor, $\alpha $ being the 
spinor index) satisfy
the `anti-periodic'  boundary condition as 
\EQ
S_{\alpha 1}(\tau, 0)=-S_{\alpha 2}(\tau, 0) , \quad 
S_{\alpha 1}(\tau, \pi)=S_{\alpha 2}(\tau, \pi) ,
\label{antiperi}
\EN
where the two-component index $i$ denotes 
the left and right moving components of the 
world-sheet spinor fields. Recall that in the case of the
standard periodic boundary  condition we would have the same 
$+$(or $-$) sign at both  end points $\sigma=0, \pi$ in
(\ref{antiperi}). The above boundary condition leads to the mass operator 
\EQ
\alpha' M^2_{GS-}= 
\sum_{n=1}^{\infty}\alpha^i_{-n}\alpha^i_n 
+ \sum_{r=1/2}^{\infty} r S_{-rA}S_{rA} -{1\over 2} ,
\EN
giving directly the above form of the partition function. 
Thus, we have established that in the GS formalism, 
two possible boundary conditions, periodic and 
anti-periodic,  of the spinor fields 
just correspond  
to the ordinary GSO projected and oppositely 
GSO projected sectors, respectively, of 
the NSR formalism. 
In this language, the fact that the closed string channel 
is automatically GSO projected is manifest, 
since the insertion of the factor $(-1)^F$ makes the 
boundary condition for the spinor field in the 
closed string channel to be periodic, irrespectively 
of boundary conditions in the open string channel. 

It is not difficult to check that the same 
mechanism which enables us to identify the  massless fermion
states with the Goldstinos corresponding to the 
spontaneous breaking of $N=2$ supersymmetry as that 
in the NSR formalism works in this formalism as well. 
Namely, the world-sheet supercharge is not conserved 
when we include the antiperiodic spinors. 
But the violation is compensated by the insertion of 
massless fermion vertex operators at open string boundaries. 
For details about this and more, I would like to invite the 
reader to the original paper \cite{yo}. 

\section{Effective low-energy theory of Goldstino}

We next briefly discuss the effective low-energy 
description of the open-string Goldstinos. One of the
characteristic features of Goldstone excitations in general 
is that the low-energy behavior of their scattering 
is constrained by various low-energy theorems reflecting 
nonlinear symmetry transformation law. In particular, 
a 4-point amplitude of them must vanish as the external 
momenta approach to zero, since the effective 
lagrangian must necessarily contain derivatives 
to preserve the symmetry under the 
transformation $\delta \psi =\epsilon + \cdots$. 
In the present case, there arises a small puzzle 
related to this: Since there exists a scalar tachyon 
$\phi$ which couples to massless fermion fields 
$\psi_{\pm}$, the indices $\pm$ being 
the chirality,   we would naively expect the appearance of the
contact term 
\[
\overline{\psi}_+\psi_- \overline{\psi}_+\psi_-
\]
in the zero-momentum limit, which is mediated by the 
exchange of the scalar tachyon  through Yukawa
interaction $\phi\overline{\psi}_+\psi_-$. 
The generic four-fermion string amplitudes indeed exhibit a 
pole corresponding to tachyon in the open-string channels
where  the opposite GSO sector can propagate, and its residue
is consistent with Yukawa coupling. But this contact term must
be cancelled  to be consistent with our interpretation of the 
massless fermion as Goldstino. It would be   a quite
nontrivial phenomenon, since only  possibility of such
cancellation is due to the  contribution of    higher massive
modes.  

Let us first derive the effective lagrangian for 
fermion scattering. A convenient way of doing this 
is to start from the generalized DBI  action
\cite{sen}\cite{schetal} (or we should 
rather call Volkov-Akulov type 
action) of a space filling
(nonBPS) D9 brane. 
\EQ
S_{{\rm  eff}}=-T \int d^{10}x 
\sqrt{-\det G_{\mu\nu}}  ,
\label{dpaction} 
\EN  
where \EQ
G_{\mu\nu}  
= \eta_{\mu\nu} + \lambda F_{\mu\nu} +
\lambda^2 S_{\mu\nu}^{(2)} + \lambda^4 S_{\mu\nu}^{(4)} ,
\EN
\EQ
S_{\mu\nu}^{(2)}= -i\overline{\psi}_+\Gamma_{\nu}\partial_{\mu}
\psi_+ - 
i\overline{\psi}_-\Gamma_{\mu}\partial_{\nu}
\psi_- ,
\EN
\EQ
S_{\mu\nu}^{(4)}=-{1\over 4}(\overline{\psi}_+\Gamma^{\alpha}
\partial_{\mu}\psi_+ \overline{\psi}\Gamma_{\alpha}\partial_{\nu}
\psi 
+
\overline{\psi}_-\Gamma^{\alpha}
\partial_{\nu}\psi_- \overline{\psi}\Gamma_{\alpha}\partial_{\mu}
\psi) . 
\EN
The ($N=2$) susy transformation law ($\psi \equiv 
\psi_{+} \oplus \psi_{-})$ is 
\EQ
\delta \psi = {1\over \lambda}\epsilon
-i{\lambda\over 2}(\overline{\epsilon}\Gamma^{\mu}\psi)
\partial_{\mu}\psi, 
\label{strpsi}
\EN
\[\hspace{-1.8cm}
\delta A_{\mu}={i\over
2}\overline{\epsilon}\Gamma_{11}\Gamma_{\mu}
\psi  +
{\lambda^2\over
24}(\overline{\epsilon}\Gamma_{11}\Gamma_{\nu}\psi
\overline{\psi}\Gamma^{\nu}\partial_{\mu}\psi
+ \overline{\epsilon}\Gamma_{\nu}\psi
\overline{\psi}\Gamma_{11}\Gamma^{\nu}\partial_{\mu}\psi )
\]
\EQ
\hspace{4cm}
-i{\lambda \over
2}(\overline{\epsilon}\Gamma^{\nu}\psi)
\partial_{\nu}A_{\mu}
-i{\lambda \over 2}(\overline{\epsilon}\Gamma^{\nu}
\partial_{\mu}\psi)
A_{\nu} .
\label{stra}
\EN
Comparing with the usual linear transformation law 
of super Yang-Mills theory,  
this form is a bit strange, since the transformation of 
$\psi$ does not contain any linear term mixing 
$\psi$ and $A_{\mu}$. However, this is an artifact 
of the choice of fields. By performing a field 
redefinition appropriately in such a way that there remains 
no cubic interaction terms in the action, 
we can rewrite the susy transformation law as 
\EQ
\delta \psi\equiv \sum_{n=-1} \lambda^n\delta \psi^{(n)} , 
\quad \delta A_{\mu}\equiv \sum_{n=0}\lambda^n \delta
A_{\mu}^{(n)}
\EN
where
\EQ
\delta \psi^{(-1)}_{\pm}=\epsilon_{\pm} ,\quad 
\delta \psi^{(0)}_\pm= \pm{1\over
4}\Gamma^{\mu\nu}\epsilon_{\pm} F_{\mu\nu} , \quad  ...
\EN
\EQ
\delta A_{\mu}^{(0)}={i\over 2}\overline{\epsilon}\Gamma_{11}
\Gamma_{\mu}\psi , \quad ...
\EN
In the lowest order with respect to derivatives, this 
reduces to the standard linear transformation law 
if we set $\psi_-=0$.  
By making expansion with respect to the power of 
fermion fields, the resulting effective action for 
4 fermion scatterings is 
\[
\hspace{-2.4cm}S_{{\rm eff}}^{(4\psi)}
=
-{\lambda^2\over 24}(\overline{\psi}_+\Gamma_{\mu}\partial_{\nu}
\psi_+\overline{\psi}_+\Gamma^{\mu}\partial^{\nu}\psi_+
+\overline{\psi}_-\Gamma_{\mu}\partial_{\nu}
\psi_-\overline{\psi}_-\Gamma^{\mu}\partial^{\nu}\psi_-
)\]
\EQ
\hspace{6cm} -{\lambda^2\over
4}\overline{\psi}_+\Gamma_{\mu}\partial_{\nu}
\psi_+\overline{\psi}_-\Gamma^{\mu}\partial^{\nu}\psi_- .
\label{expandedaction2}
\EN
Thus, the 4 fermion amplitudes must vanish as the 
squares of momenta. We have confirmed 
that the required
cancellation indeed occurs between the tachyon exchange 
and that of higher excitation modes of open string 
in the oppositely GSO projected sector. For this, 
the $s$-$t$ duality symmetry 
\[
A_{1+,3-,2+,4-}(s,t)=A_{1+,4-,2+,3-}(t,s)
\]
between the mutually dual planar amplitudes with 
cyclic ordering $(1+, 3-, 2+,4-)$ is crucial: 
This symmetry is in contradiction with the 
fermi antisymmetry with respect to the 
exchange $1 \leftrightarrow 2$, and hence this particular 
ordering does not contribute to  the final amplitude. 
The peculiarity occurring here is related to the apparent
violation  of spin-statistics theorem in the NSR formalism
when  both the GSO sectors are included. 
The zero-momentum limit of this 
planar amplitude does not vanish, while the other planar 
contributions with orderings 
such as $(1+, 2+, 3-, 4-)$ contribute, but do vanish in the
zero-momentum limit. Combining all these properties, it turns
out that the net results for the  low-energy behavior of the
total  string amplitudes agree precisely with the 
prediction of the above low-energy effective action. 
The indispensable role of the higher excited states and 
open-open string duality 
is consistent with our previous observation 
on the importance of the open-closed string duality. 
For the fuller account of the above remarkable 
properties of Goldstino scatterings in 
superstring theory, I would like to refer the reader to our 
paper \cite{harayo}. 
 
\section{World-sheet mechanism for susy restoration ?}

As the final topic of this talk, I would like to speculate on
the possible mechanism of restoring supersymmetry from its 
spontaneously broken phase.  In the low-energy
approximation of the previous section,  the supersymmetry is
restored if we have the condensation of fermion bilinears 
characterized by 
$\langle
i\overline{\psi}_{\pm \beta}\partial_{\mu}
\psi_{\pm\alpha}\rangle
=-(\Gamma_{\mu}(1\pm \Gamma_{11})/2)
_{\alpha\beta}/5\lambda^2$,  in terms of the fermion fields
before the  field redefinition which cancels the 
inhomogeneous term of the susy transformation law. 
If we wish to avoid the low-energy approximation, an obvious 
approach to this question would be to use  string field
theory. In fact, there is a trivial way  of eliminating the inhomogeneous term in
(\ref{fermitrans}), namely the shift 
$a \rightarrow a - Y({\pi\over 2})Q_I {\cal I} $. 
This completely eliminates the kinetic term and 
makes the resulting action purely cubic. 
However, this shift of string field is not yet the
condensation of tachyon,  since the shifted part actually
does {\it not}  correspond to the oppositely GSO projected
sector.  To obtain the stable open string vacuum, we would have
to further shift the opposite GSO sector by an
appropriate classical  string field configuration which is
annihilated by  the supercharge $Q_{\alpha}$. 

Actually, even before going to this
problem, the trivial shift of this kind raises a puzzle. The
value of  the action at the classical vacua is zero either
before or after the above shift of string field. But, the
original action before any shift is supposed to describe an
unstable D9 brane. Then, after restoring supersymmetry,  the
vacuum value of the action must have been lowered by
the  tension of the unstable D9 brane.  This puzzle seems to
be indicative of  the danger involved in the formal
manipulations  such as the ones  associated with the identity
field, midpoint insertion  of picture changing operator, and
so on. With regard to these difficulties,  a promising
possibility may be to use the version  of superstring theory
proposed in
\cite{berko}, as  already mentioned before.  

Instead of pursuing this problem further within 
the framework of string field theory, 
let us turn our discussion to the world-sheet 
approach. Our intuitions on various dynamical 
questions such as symmetry breaking and its 
restoration usually depend  
on field theory. However, one of the
remarkable  features of string theory is that  
physical pictures on various field theoretical  
phenomena are remarkably simplified often by using 
rather the naive world-sheet picture. 
It would be very nice if we have such an example 
again in the present problem. In fact, the so-called
boundary string  field approach \cite{bsft}  clearly suggests
a simple physical picture that  after tachyon condensation
only surviving open strings  are those with the world sheets
whose boundaries shrink to  points in target  space-time. Then
it is natural  to regard the open string boundaries as some
sort of defects on closed-string world sheets. From such a
viewpoint,  the restoration of $N=2$ susy could occur when 
the boundary conditions for the world-sheet  spinors were 
dynamically flipped at the defects  such that the left and
right movers  propagate independently of each other. 

The questions are then how such a flip of boundary condition 
can occur and how to treat such a phenomenon dynamically. 
Kinematically, it seems clear that 
there must be the same number of dynamical degrees 
before and after the flip at the  defect. 
Suppose that the defect is located between
$\sigma=\sigma_0$ and $\sigma_0+a$, 
$a$ being the lattice constant. If the defect corresponds 
to an unstable D-brane or to a D-brane and anti-D brane
system,  there are open strings with all possible combinations 
of the boundary 
conditions, 
\[
S_+(\sigma_0+a)=0 \quad {\mbox or} \quad S_-(\sigma_0+a)=0
\]
and 
\[
S_+(\sigma_0)=0 \quad {\mbox or} \quad S_-(\sigma_0)=0 ,
\]
where we defined new combinations of spinor components
 $S_{\pm}=(S_1\pm S_2)/2$. 
For instance, in the case of D9-${\overline{{\mbox D}9}}$ 
system in type IIB theory, we can consistently choose 
$S_+=0$ and $S_-=0$ at the boundaries attached to 
D-brane and anti-D-brane, respectively. The 
boundary conditions which preserve one half 
of supersymmetry are those with the same 
$\pm$ indices (viz. DD and $\overline{{\rm D}}
\overline{{\rm D}}$ strings) at the both
boundaries, while  the nonsupersymmetric ones 
(viz. D$\overline{{\rm D}}$ and 
$\overline{{\rm D}}$D strings) correspond to the choices  
of different signs at the two end points.  The closed string 
configuration after the annihilation of the defect on 
the other  hand is represented as 
$S_{\pm}(\sigma_0+a)=
S_{\pm}(\sigma_0)$, realizing $N=2$ susy provided that the 
annihilation of the  defect occurred also in the bosonic part. 
The possibility of dynamically flipping from the 
open-string boundary conditions to 
that of closed string 
arises only when open string excitations with 
all the above combinations are included, since then 
appropriate recombinations of spin components can in 
principle lead to the flip.  
Contrary to this, 
in the case of BPS  D-branes, either
one of $S_+$ or $S_-$ vanishes at all times and hence 
we can never excite the spinor components such that
  these two
components $S_{\pm}$ become independent  propagating modes. 

Of course, our argument is very naive 
and is not sufficient to  show that the flip of boundary
condition of the desired type can be achieved.  For that, we
need to develop a systematic  formulation of D-branes and   
fundamental strings such that the creation and  annihilation
of D-branes and associated change of boundary conditions  can
be dealt with  as {\it dynamical} processes on closed-string
world sheets.   One candidate for such possibilities is
supermembrane theory, or matrix-string theory 
as its regularized version. It is possible to interpret 
 D0($\overline{\mbox{D0}}$)-branes as 
soliton-like excitations on supermembranes. In
connection with this,  we have recently proposed \cite{sekyo}
a systematic method of directly mapping supermembranes 
wrapped around a circle  to matrix strings in the large 
$N$ limit. Recall that matrix-string theory 
is a framework in which both fundamental strings and 
D-particles can be treated dynamically. I hope
that this new development will be  useful  for discussing
various dynamical issues related  to the hidden susy of
string/M theory and other important aspects 
of nonperturbative string theory.

%\begin{figure}
%  \resizebox{6pc}{!}{\includegraphics{test}}
%  \resizebox{6pc}{!}{\reflectbox{\includegraphics{test}}}
%\caption{The caption}
%\end{figure}

\vspace{1cm}
\noindent
Acknowledgments

\vspace{0.5cm}
I would like to thank N. Berkovitz for collaborative 
discussions about the possibility of realizing the  hidden 
susy using his approach to superstring field theory. 
I am also grateful to Y. Matsuo for stimulating 
discussions on string field theory during the symposium. 
The present work is supported in part by Grant-in-Aid for Scientific 
Research (No. 12440060)  from the Ministry of  Education,
Science and Culture.

\end{document}